# Economic Diversification and Social Progress in the GCC Countries: A Study on the Transition from Oil-Dependency to Knowledge-Based Economies

**Mahdi Goldani**

m.goldani@hsu.ac.ir

**Soraya Asadi**

S_asadi@atu.ac.cr

Abstract:

The Gulf Cooperation Council countries -Oman, Bahrain, Kuwait, UAE, Qatar, and Saudi Arabia- holds strategic significance due to its large oil reserves. However, these nations face considerable challenges in shifting from oil-dependent economies to more diversified, knowledge-based systems. This study examines the progress of Gulf Cooperation Council (GCC) countries in achieving economic diversification and social development, focusing on the Social Progress Index (SPI), which provides a broader measure of societal well-being beyond just economic growth. Using data from the World Bank, covering 2010 to 2023, the study employs the XGBoost machine learning model to forecast SPI values for the period of 2024 to 2026. Key components of the methodology include data preprocessing, feature selection, and the simulation of independent variables through ARIMA modeling. The results highlight significant improvements in education, healthcare, and women's rights, contributing to enhanced SPI performance across the GCC countries. However, notable challenges persist in areas like personal rights and inclusivity. The study further indicates that despite economic setbacks caused by global disruptions, including the COVID-19 pandemic and oil price volatility, GCC nations are expected to see steady improvements in their SPI scores through 2027. These findings underscore the critical importance of economic diversification, investment in human capital, and ongoing social reforms to reduce dependence on hydrocarbons and build knowledge-driven economies. This research offers valuable insights for policymakers aiming to strengthen both social and economic resilience in the region while advancing long-term sustainable development goals.

## 1. Introduction

The GCC countries region is one crucial area that contains an estimated 50 percent of the world's oil reserves and thus plays a strategic geopolitical role. Due to the specific location of this region, it has given a significant economic and strategic position. the countries that this region is encompassed Oman, Bahrain, Kuwait, the United Arab Emirates, Qatar, and Saudi Arabia. In addition to the rich natural resources of this region, they also showcase its significant cultural heritage and economic potential. Despite the progress potential of this region, the progress path of the countries has not always been smooth and has faced huge challenges. These challenges can be discussed in economic, political and environmental fields.

The main economic progress challenge refers to the oil-based economy. Oil was discovered in GCC countries aria and by rising world demand, revenue of oil exports mounted. Oil and associated activities came to dominate countries' economic structures [1]. Overcoming the major challenges hinder the transition to knowledge-based economies implies changing the economic structure, by shifting from natural resources-oil-based economies (rent-seeking economies) to knowledge-based economies in Arab Gulf countries. [2] Transitioning from an oil-based economy to a more diversified one is challenging for these economic systems that require significant investment in new industries, development of human capital, and fostering an entrepreneurial ecosystem. The members of the Gulf Cooperation Council GCC face a difficult problem in refocusing growth paradigms in order to diversify their economies. However, doing so will minimize their dependency on hydrocarbons, empower the private sector to drive growth, and provide people with the skills needed to enter the high-value-added jobs created by those economies [3]. During the transition from oil economy and globalization, the link between economy and politics is very important [4]. Social and political reforms are critical elements for fostering progress in any region, and the GCC countries are no exception. These reforms are essential for creating more inclusive, equitable, and stable societies, which in turn can drive economic growth and development. By investing in education, healthcare, women's rights, governance, and political participation, these nations can create more inclusive, stable, and prosperous societies. The real non-hydrocarbon economy continues to perform strongly, seeing uninterrupted growth as a full recovery from dissipating COVID-19 pandemic-related uncertainties comes into play.

fig1. GCC: Hydrocarbon/Non-hydrocarbon contribution to real GDP growth

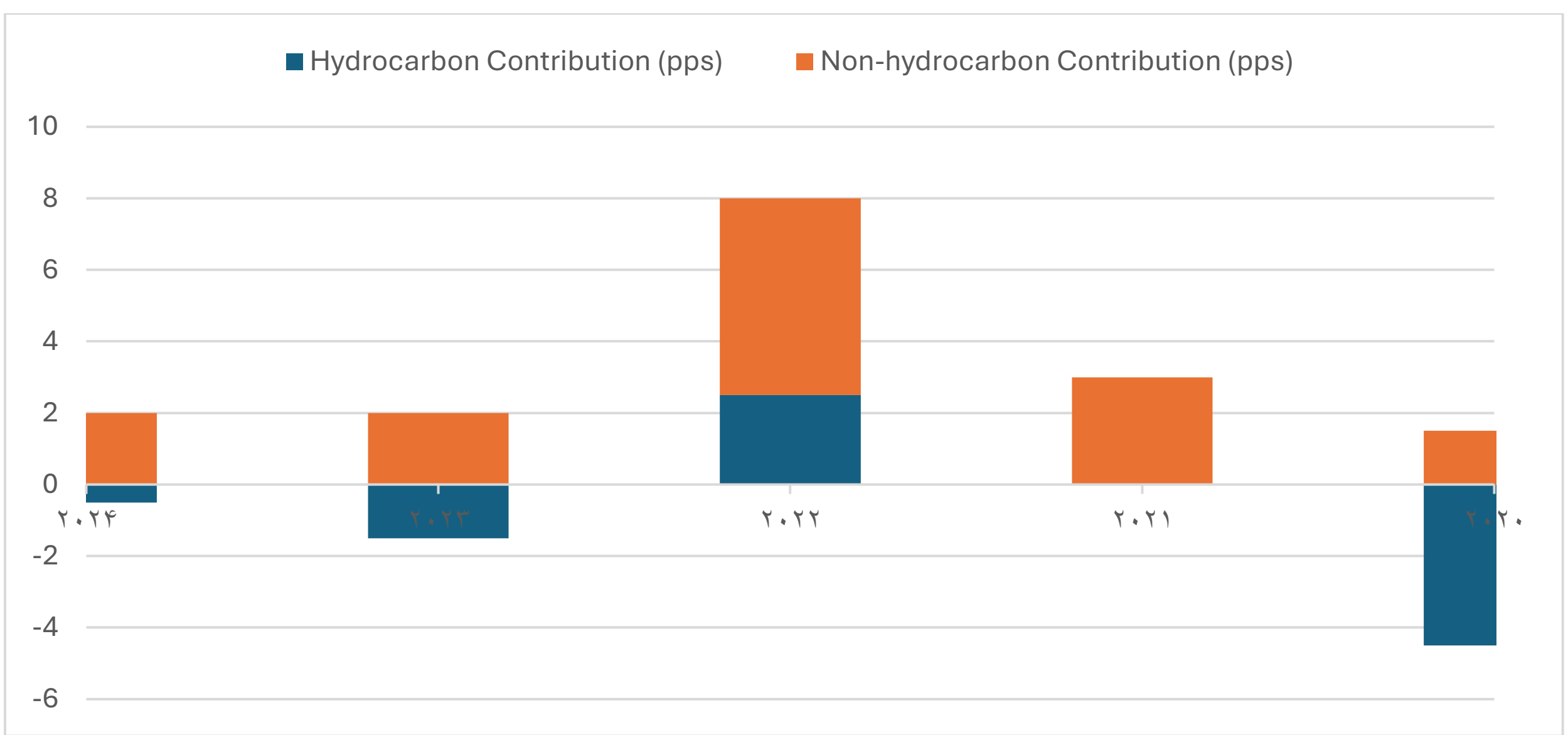

Source: S&P Global Market Intelligence, 2023

To get comprehensive overview of progress issue GCC countries, the progress to achieve the Sustainable Development Goals (SDG) can be examined. in All the GCC countries are progressing with the SDGs' agenda implementation and four of the countries have already exceeded two-thirds of the overall SDG score (Oman, Qatar, Saudi Arabia and UAE), with the UAE and Saudi Arabia fully achieving three and two SDGs, respectively. The improvement in the SDG score was mainly driven by more seats for women in parliament in Bahrain, Kuwait, Oman, and the UAE, higher manufacturing participation in GDP and employment across all countries, enhanced access to technology, increased share of salaries in GDP in Kuwait and Oman, improved statistical performance and higher government spending on health and education [5].

Fig2. Sustainable Development Goals

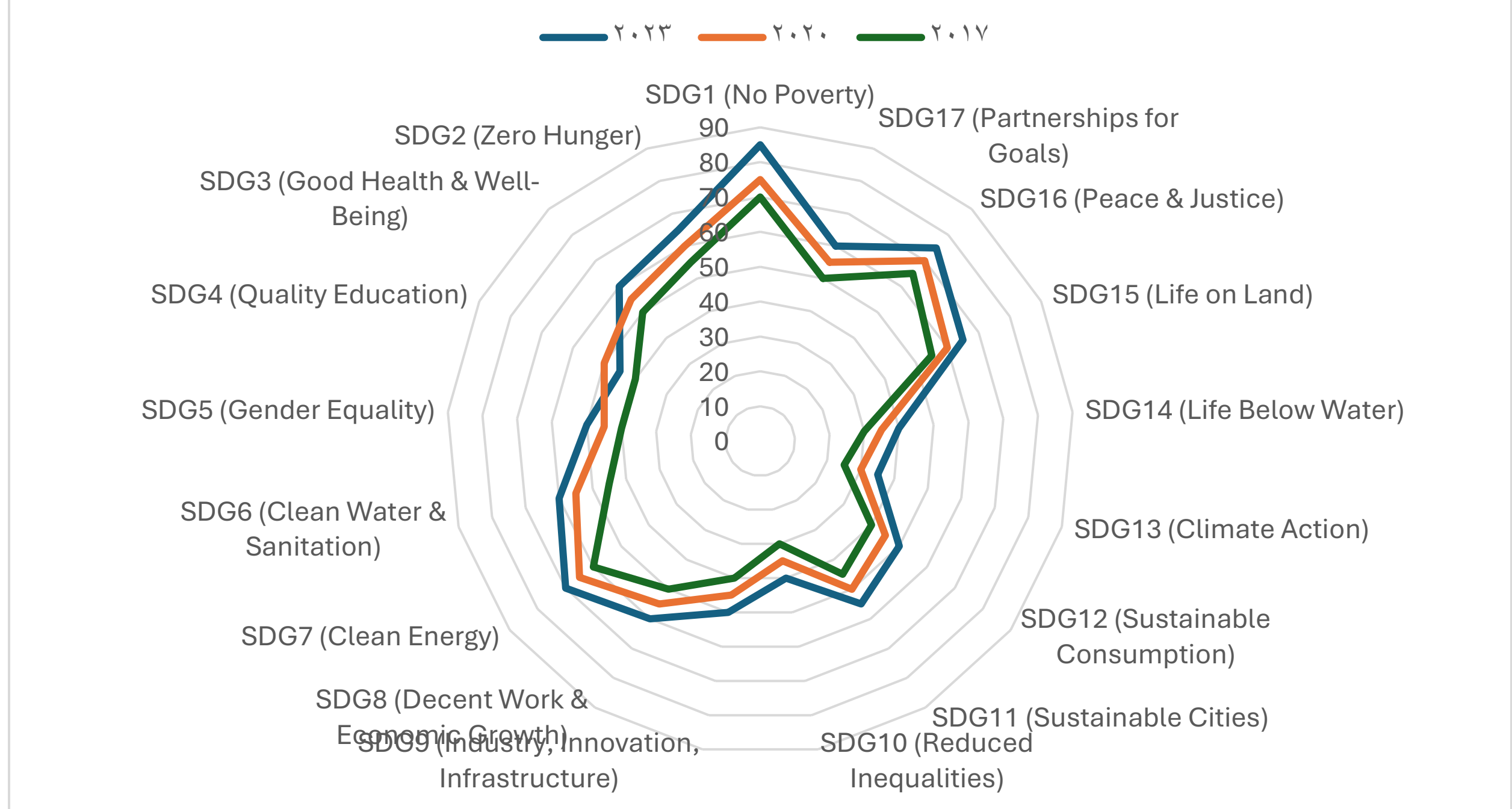

Source: Atlas of Sustainable Development Goals 2023

To measure the level of progress in a society, various criteria are embedded, each of which measures a different aspect of progress. While some authors believe the focusing on living standards don't provide a comprehensive view of social progress [6], Social Progress Index (SPI) which proposed by Porter [7] provide more comprehensive picture of progress because it includes a wider range of social and environmental indicators that directly impact human well-being, beyond just economic measures. Comparing the SPI in the GCC countries shows While Kuwait and UAE lead in the region with higher SPI rankings, all GCC countries share common challenges, particularly in the domains of personal rights and inclusiveness. The strengths in basic human needs and foundations of wellbeing are consistent, but the opportunity dimension remains an area needing significant attention Despite the challenges faced by GCC countries, the Social Progress Index (SPI) indicates notable progress in these nations [8].

fig3. social progress index

Source: The Global Social Progress Index, 2023

The economic crisis and decline in oil prices caused by the pandemic led governments in the Gulf Cooperation Council (GCC) region to step up their diversification efforts [9]. The aim of GCC is to reduce dependence on oil revenues [10], in order to achieve this goal, economic diversification is the main strategy of the Gulf Cooperation Council (GCC) [11]. They attempt to achieve diversification by building a knowledge-based economy, focusing on the quality of education and research that improves the human capital available in the country which contributes to the growth of the economy [12]. continuous efforts to develop the status of ICT, education, innovation, and entrepreneurship in several GCC countries have contributed to improving their international competitiveness, as seen by advancements in rankings issued by various international organizations [9]. In the GCC, knowledge-based economy results from top-down economic policies that promote entrepreneurship, R&D, environment, and innovation while also reforming the education system [13, 14]. in addition, the GCC countries have been investing in improving the knowledge, skill, health of their population, and join the world bank human capital project. These countries score on human capital index are, however, lower than those countries at comparable level of income [15], however, statistics show that there has been good progress in the field of human capital in recent years in these countries [16].

## 2. Methodology

In this section the structure of the model is examined. Data collection is the first step. As the research aims to predict future values of SPI in GCC, a comprehensive dataset is needed. A comprehensive dataset is a dataset that includes variables that represent the economic, social, health, education, and political status in order to provide a proper analysis of the future of the target variable. For this purpose, world bank development indicators are proper. The database contains 1,400 time series indicators for 217 economies and more than 40 country groups, with data for many indicators going back more than 50 years. These variables were extracted for six countries-Oman, Bahrain, Kuwait, the United Arab Emirates, Qatar, and Saudi Arabia. The dataset encompasses the period from 2010 to 2023, offering a robust temporal range for analysis. This period was selected to capture recent trends and provide sufficient historical context for accurate forecasting. The figure1 show the methodology process of this research.

Fig4. the methodology process of this research

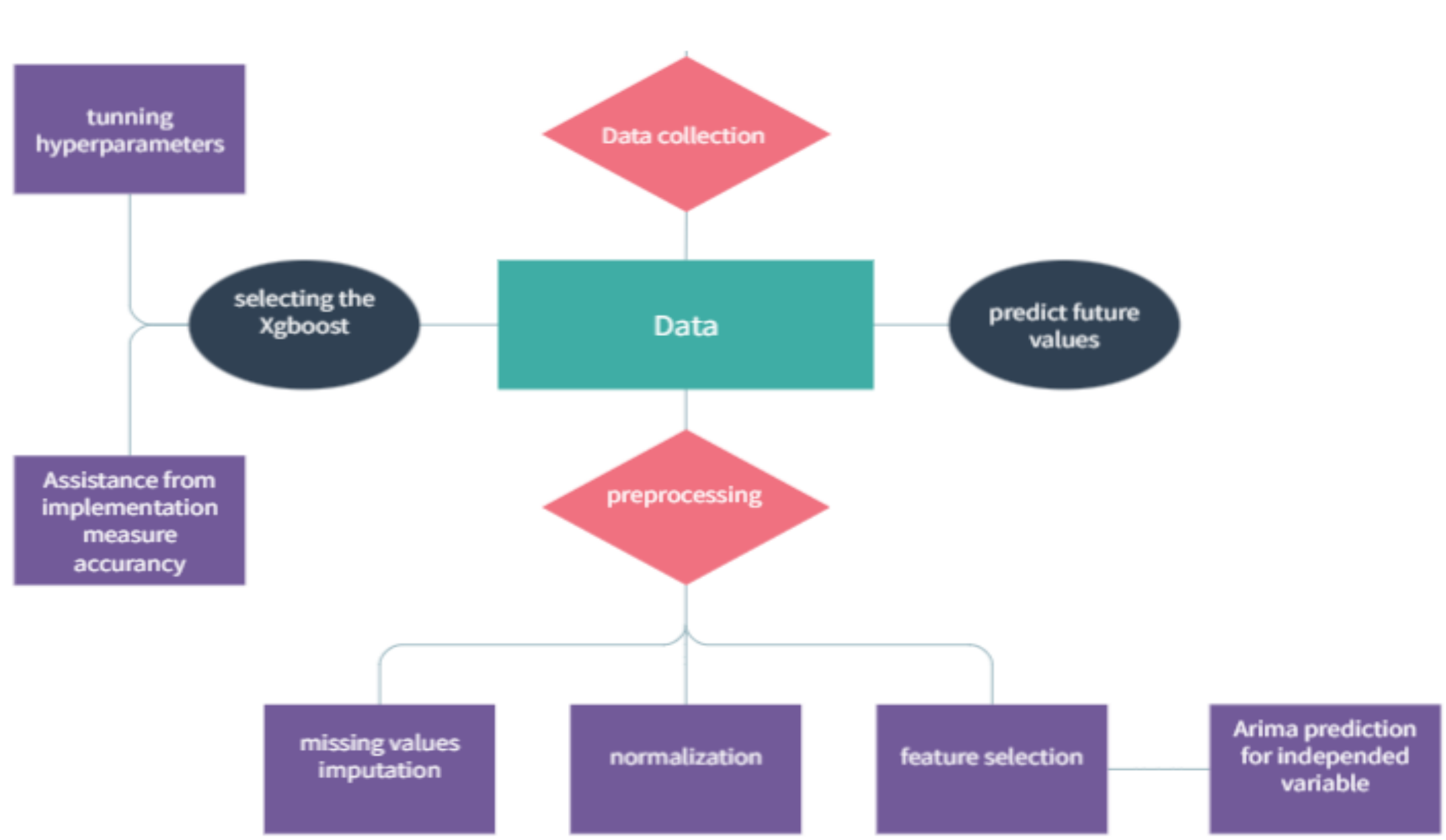

## 2-1. Data Preprocessing

One of the primary challenges in working with real-world data is the presence of missing values. In our dataset, variables with more than 70% missing data were discarded to maintain data quality and integrity. For the remaining data, missing values were imputed using the random forest method. This method was chosen for its high accuracy and ability to handle complex data structures, as evidenced by Goldani [17].

## 2-2. Feature Selection

Given the high-dimensional nature of the World Bank Development Indicators, feature selection was a critical step to mitigate issues such as overfitting, computational complexity, and noise sensitivity. The Edit Distance on Real sequence (EDR) method was chosen for feature selection due to its lower sensitivity to sample size and simpler calculations [18]. Based on the target variable, the social progress index, table 1 shows 8 of the most important factors affecting this index that were selected from each dataset.

Table1. selected features in datasets

| | Bahrain | UAE | Kuwait | Qatar | Saudi Arabia | Oman |
|---|---|---|---|---|---|---|
| 1 | GDP per capita growth (annual %) | Adjusted savings: net forest depletion (current US$) | Arms imports (SIPRI trend indicator values) | Annual freshwater withdrawals, domestic (% of total freshwater withdrawal) | Adjusted savings: net forest depletion (current US$) | Energy use (kg of oil equivalent per capita) |
| 2 | Life expectancy at birth, total (years) | Claims on central government (annual growth as % of broad money) | Computer, communications and other services (% of commercial service exports) | Children out of school, primary, male | Cereal production (metric tons) | Fertility rate, total (births per woman) |
| 3 | Merchandise exports by the reporting economy, residual (% of total merchandise exports) | CO2 emissions from gaseous fuel consumption (% of total) | Current account balance (% of GDP) | CO2 intensity (kg per kg of oil equivalent energy use) | CO2 emissions from liquid fuel consumption (kt) | Fertilizer consumption (kilograms per hectare of arable land) |
| 4 | Population ages 0-14, female (% of female population) | CO2 emissions from liquid fuel consumption (% of total) | Gross intake ratio in first grade of primary education, total (% of relevant age group) | Mortality rate, under-5 (per 1,000 live births) | Food exports (% of merchandise exports) | Gross fixed capital formation (% of GDP) |
| 5 | School enrollment, primary (gross), gender parity index (GPI) | Fertilizer consumption (% of fertilizer production) | Land under cereal production (hectares) | Population ages 10-14, male (% of male population) | GDP growth (annual %) | Imports of goods and services (% of GDP) |
| 6 | Transport services (% of service imports, BoP) | Merchandise exports to economies in the Arab World (% of total merchandise exports) | Machinery and transport equipment (% of value added in manufacturing) | Primary education, pupils (% female) | Industry (including construction), value added (constant 2015 US$) | Natural gas rents (% of GDP) |
| 7 | Service exports (BoP, current US$) | Urban population (% of total population) | Net primary income (Net income from abroad) (current US$) | Pupil-teacher ratio, tertiary | Merchandise exports to low- and middle-income economies in South Asia (% of total merchandise exports) | Ores and metals imports (% of merchandise imports) |
| 8 | Travel services (% of service exports, BoP) | Water productivity, total (constant 2015 US$ GDP per cubic meter of total freshwater withdrawal) | Portfolio investment, net (BoP, current US$) | Survival to age 65, male (% of cohort) | Population ages 15-64, total | Population ages 00-04, female (% of female population) |

### 2-3. Simulation of Independent Variables

To predict the social progress index for the years 2024 to 2026, it was necessary to simulate the independent variables for these years. The ARIMA (AutoRegressive Integrated Moving Average) method was used for this purpose, given its effectiveness in time series forecasting [19]. The final features were chosen based on two criteria: the appropriate length of the time series and suitable fittings for all ARIMA models. Variables that did not meet these criteria were excluded to ensure the robustness of the predictive model.

### 2-4. XGBoost model

The XGBoost machine learning method was selected for this study due to its superior performance characteristics, including high accuracy, speed, robustness to overfitting, and flexibility [20]. XGBoost, or Extreme Gradient Boosting, is an optimized distributed gradient boosting library designed to be highly efficient, flexible, and portable. It has become a favored choice for various machine learning competitions and practical applications because of its scalability and performance.

Table2. Comparison of XGBoost with Other Prediction Methods in Machine Learning

| Method | Strengths | Weaknesses | Best Use Cases |
| --- | --- | --- | --- |
| XGBoost | High accuracy, speed, handles large datasets, built-in regularization to prevent overfitting | More complex to tune, computationally intensive for large hyperparameter grids | Large datasets, competitions, handling missing data, high performance required |
| Random Forest | Good for preventing overfitting, easy to interpret, works well with categorical data | Less efficient on very large datasets, less accurate than boosting algorithms | Smaller or medium-sized datasets, when interpretability is important |
| Support Vector Machines (SVM) | Effective for complex decision boundaries, works well on smaller datasets | Computationally expensive for large datasets, sensitive to feature scaling | Smaller datasets, complex classification problems, high-dimensional spaces |
| Neural Networks | Excels with unstructured data, can model complex relationships | Prone to overfitting, requires large datasets, complex to tune | Unstructured data (e.g., images, text), when deep feature extraction is needed |

Source: [17]

Gradient boosting is a machine learning technique for regression and classification problems, which builds a model in a stage-wise fashion by sequentially adding predictors to an ensemble, each correcting its predecessor's errors. XGBoost leverages this framework to significantly improve model performance. By focusing on the residuals of previous models, it refines its predictions iteratively, which enhances accuracy. XGBoost incorporates L1 (Lasso) and L2 (Ridge) regularization techniques to prevent overfitting. These techniques add penalties to the model for having too many features or coefficients with large magnitudes, thus helping in creating a robust model that generalizes well to unseen data. Regularization is essential for controlling the model's complexity and preventing it from fitting the noise in the training data [21]. One of the key advantages of XGBoost is its support for parallel processing. It can utilize multiple CPU cores to perform computations, significantly speeding up the training process. This is achieved by constructing trees in parallel during the boosting process, making XGBoost much faster compared to other boosting algorithms [22]. XGBoost can handle missing data internally by learning the best imputation values during the training process. This feature is particularly useful when working with real-world datasets that often contain missing values. By automatically handling missing values, XGBoost simplifies the data preprocessing steps and ensures that the model is not biased by imputed data [23]. XGBoost uses a depth-first approach to tree pruning, employing a heuristic to ensure that the trees do not grow too complex. This is accomplished by setting a maximum depth for the trees and using a minimum loss reduction threshold for further splitting. Tree pruning helps in controlling overfitting by limiting the complexity of the model, thereby ensuring better generalization to new data [21]. The algorithm can handle sparse data efficiently, which is useful in datasets with many missing values or zero entries. XGBoost is designed to be aware of sparsity patterns in the data and optimizes the computation accordingly. This feature is particularly beneficial for large-scale machine learning problems where data sparsity is a common issue [24]. XGBoost allows users to define their own objective functions and evaluation metrics, providing flexibility to tailor the model to specific needs. This customization enables the model to be adapted for various applications, ranging from regression to classification and ranking problems [23]. To find the optimal combination of these hyperparameters, the grid search method was employed. Grid search systematically works through multiple combinations of parameter values, cross validating each combination to determine which provides the best performance. This method helps in identifying the best hyperparameters that maximize the model's predictive accuracy while minimizing overfitting [25].

**2-5. Implementation of Grid Search**

The grid search method involves setting up a parameter grid with possible values for each hyperparameter and then evaluating the model's performance for each combination through cross-validation. The hyperparameters tuned in this research included:

- n_estimators: The number of trees in the ensemble. A higher number of trees increases model complexity but also the risk of overfitting.
- learning_rate: Also known as 'eta', this parameter controls the step size shrinkage used to update the weights. A lower learning rate makes the model more robust to overfitting but requires more trees.
- max_depth: The maximum depth of each tree. Increasing the depth makes the model more complex, allowing it to capture more details from the training data, but it also increases the risk of overfitting.
- subsample: The fraction of samples used to fit each individual tree. Lowering the subsample value can help prevent overfitting by introducing randomness into the training process.

By selecting XGBoost and meticulously tuning its hyperparameters, the study ensured the development of a robust and accurate model capable of effectively predicting the social progress index. The grid search method played a crucial role in optimizing the model, balancing the complexity and performance to avoid overfitting while maintaining high predictive accuracy.

**2-6. Model Evaluation**

The dataset was split into training and testing sets. Due to the limited sample size, only two rows were reserved for testing, with the remainder used for training. This approach ensures that the model is evaluated on unseen data to check its generalizability.

The mean absolute percentage error (MAPE) was employed to assess the prediction accuracy for both training and testing data. MAPE is calculated as follows:

$$\text{MAPE} = \frac{1}{n}\sum_{i=1}^{n}\left|\frac{A_i - F_i}{A_i}\right| \times 100$$

where$A_i$ is the actual value and $F_i$ is the forecasted value. This metric provides a straightforward and interpretable measure of forecast accuracy [23].

## 3. Result

The first step in the prediction process involves data preprocessing, which includes handling missing values, normalizing the data, and selecting relevant features. After collecting data from the World Bank Development Indicators for six countries (Oman, Bahrain, Kuwait, the United Arab Emirates, and Saudi Arabia), the missing values were addressed. Variables with more than 70 percent missing data were discarded. Subsequently, missing values were imputed using the random forest method, recognized for its accuracy [17]. The World Bank Development Indicators are high-dimensional, which can lead to issues such as overfitting, computational complexity, and noise sensitivity. Therefore, selecting the most important variables was the initial step of this research. Given the data size limitations, precision in feature selection methods for small datasets was crucial. Two main factors were considered: the high accuracy of methods in small datasets and simplicity in calculations. The EDR (Edit Distance on Real sequence) method was chosen for feature selection due to its lower sensitivity to sample size and simpler calculations [18].

All development indicators from the World Bank were extracted for the six countries, covering the period from 2010 to 2023. Based on the target variable, the social progress index, twenty of the most relevant indicators were selected from each dataset. To predict the social progress index for 2024 to 2026, the independent variables for these years needed to be simulated using the ARIMA method. From the twenty selected variables, the final features were chosen based on two criteria: the appropriate length of the time series and suitable fittings for all ARIMA models. Variables that did not meet the sufficient length criterion were removed, followed by the removal of ARIMA models with accuracy less than 80%. The remaining variables (as listed in Table 3) were used as main characteristics to predict the social progress index in each dataset.

Table3. The independent variables of datasets

| **United Arab Emirate** | **Oman** | **Bahrain** | **Kuwait** | **Qatar** | **Saudi Arabia** |
|---|---|---|---|---|---|
| Adjusted savings: net forest depletion (current US$) | Energy use (kg of oil equivalent per capita) | GDP per capita growth (annual %) | Computer, communications and other services (% of commercial service exports) | Annual freshwater withdrawals, domestic (% of total freshwater withdrawal) | Adjusted savings: net forest depletion (current US$) |
| Claims on central government (annual growth as % of broad money) | Fertility rate, total (births per woman) | Life expectancy at birth, total (years) | Land under cereal production (hectares) | Children out of school, primary, male | Cereal production (metric tons) |
| CO2 emissions from gaseous fuel consumption (% of total) | Fertilizer consumption (kilograms per hectare of arable land) | Merchandise exports by the reporting economy, residual (% of total merchandise exports) | Machinery and transport equipment (% of value added in manufacturing) | CO2 intensity (kg per kg of oil equivalent energy use) | CO2 emissions from liquid fuel consumption (kt) |
| CO2 emissions from liquid fuel consumption (% of total) | Gross fixed capital formation (% of GDP) | Population ages 0-14, female (% of female population) | Net primary income (Net income from abroad) (current US$) | Mortality rate, under-5 (per 1,000 live births) | Food exports (% of merchandise exports) |
| Fertilizer consumption (% of fertilizer production) | Imports of goods and services (% of GDP) | School enrollment, primary (gross), gender parity index (GPI) | Portfolio investment, net (BoP, current US$) | Population ages 10-14, male (% of male population) | GDP growth (annual %) |
| Merchandise exports to economies in the Arab World (% of total merchandise exports) | Natural gas rents (% of GDP) | Transport services (% of service imports, BoP) | Primary income payments (BoP, current US$) | Primary education, pupils (% female) | Industry (including construction), value added (constant 2015 US$) |
| Urban population (% of total population) | Ores and metals imports (% of merchandise imports) | Service exports (BoP, current US$) | Secondary education, teachers, female | Pupil-teacher ratio, tertiary | Merchandise exports to low- and middle-income economies in South Asia (% of total merchandise exports) |
| Water productivity, total (constant 2015 US$ GDP per cubic meter of total freshwater withdrawal) | Population ages 00-04, female (% of female population) | Travel services (% of service exports, BoP) | Secondary education, vocational pupils | Survival to age 65, male (% of cohort) | Population ages 15-64, total |
| | School enrollment, primary and secondary (gross), gender | | | | |

| | parity index (GPI) |
|---|---|

After preparing the six datasets for this study, the XGBoost machine learning method was selected due to its high accuracy, speed, robustness to overfitting, and flexibility [18]. Before implementing the model, the hyperparameters were optimized using the grid search method. The XGBoost hyperparameters tuned in this research included 'n_estimators', 'learning_rate', 'max_depth', and 'subsample' (as detailed in Table 4).

Table4. hyperparameters of XGboost model

| **HYPERPARAMETERS** | **UNITED ARAB EMIRATE** | **OMAN** | **BAHRAIN** | **KUWAIT** | **QATAR** | **SAUDI ARABIA** |
|---|---|---|---|---|---|---|
| **LEARNING_RATE** | 0.01 | 0.2 | 0.2 | 0.2 | 0.2 | 0.1 |
| **MAX_DEPTH** | 3 | 6 | 6 | 3 | 3 | 6 |
| **N_ESTIMATORS** | 100 | 200 | 100 | 200 | 200 | 200 |
| **SUBSAMPLE** | 1 | 0.9 | 1 | 0.9 | 0.8 | 1 |

After tuning the hyperparameters, the XGBoost model was evaluated. The first step in the forecasting process involved splitting the data into training and testing sets. Due to the limited sample size, only two rows were reserved for testing, with the remainder used for training. The mean absolute percentage error (MAPE) was employed to assess the prediction accuracy for both training and testing data. This metric provides a straightforward and easily interpretable measure of forecast accuracy [26]. As shown in Figure 1, the training MAPE values tend to be lower, indicating that the model has been optimized to fit the training data closely. In four out of the six datasets, the training MAPE was nearly zero, suggesting potential overfitting, where the model performs exceptionally well on the training data but may not generalize well to new, unseen data. To address this issue, the MAPE for the testing data was also calculated (Figure 2).
The performance of the model varied across the datasets, with the best results observed for Bahrain and Kuwait. Overall, the forecast performance was favorable, as all MAPEs were below ten percent, indicating good predictive accuracy.

Fig6. MAPE of training data

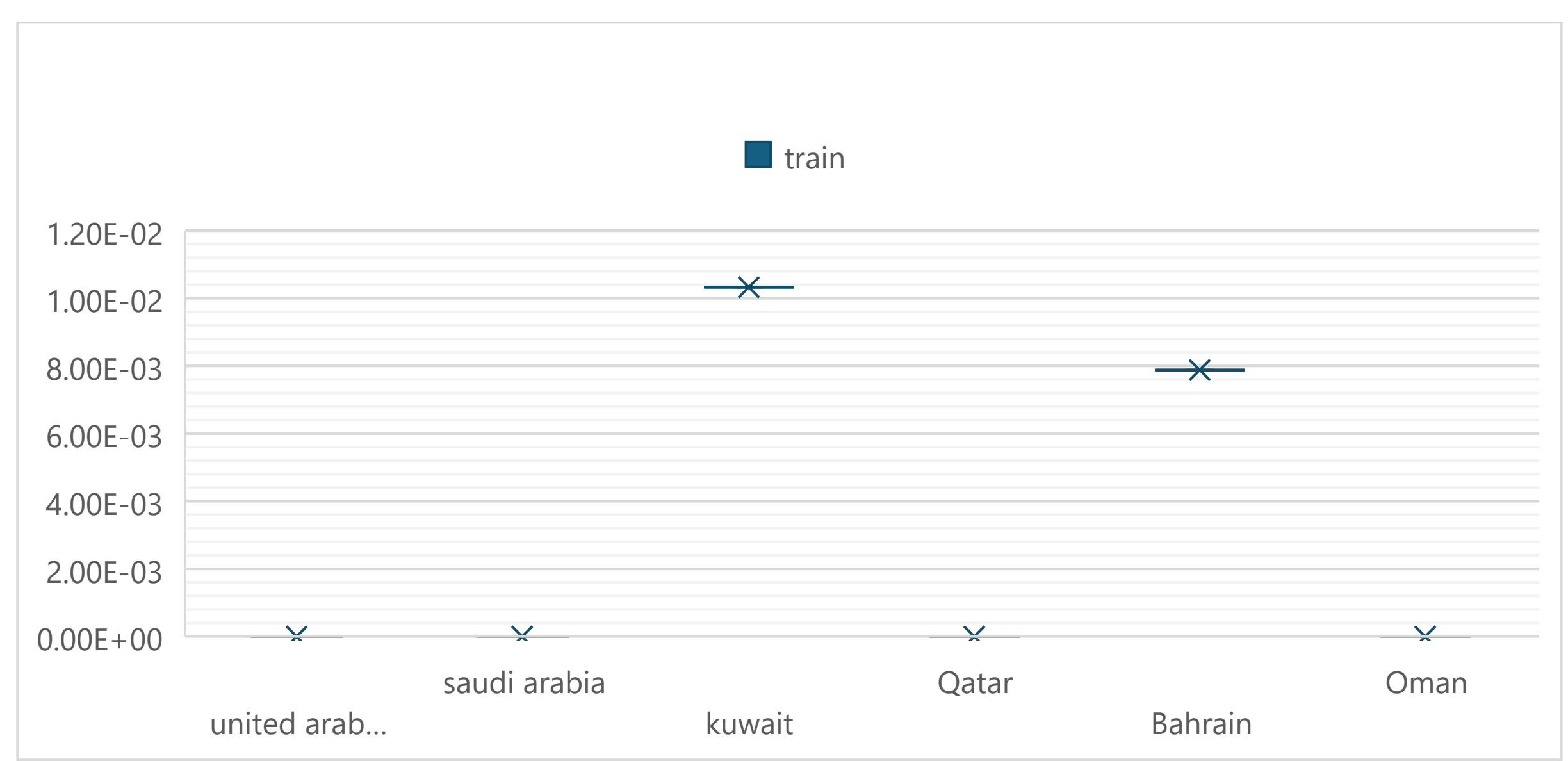

Fig7. MAPE of testing data

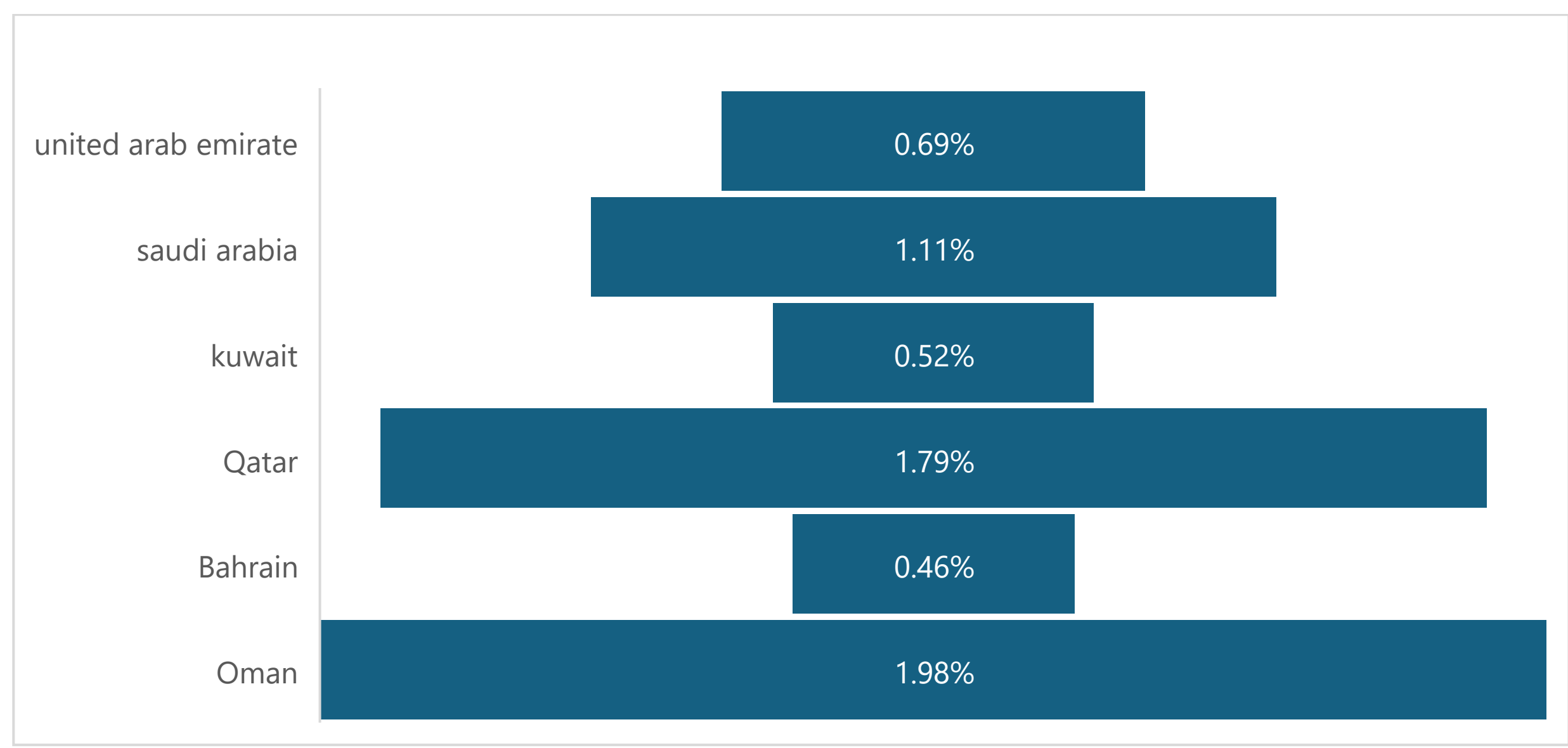

By considering the ARIMA prediction for each feature, the forecast was made for the years 2027-2023 for all datasets. The trend of changes in the social progress index until 2022 has been positive for all six countries under review. The trend of changes in the index of social progress until 2020 has been positive for all six countries. After that, as shown in Figure 2, all six countries witnessed a decrease in the value of this index, and in some countries, this decrease occurred with a steeper slope. Test data for all graphs show that this method predicts the reduction well.

Fig8. Actual vs Predicted Social Progress Index (SPI) for GCC Nations

UAE Bahrain

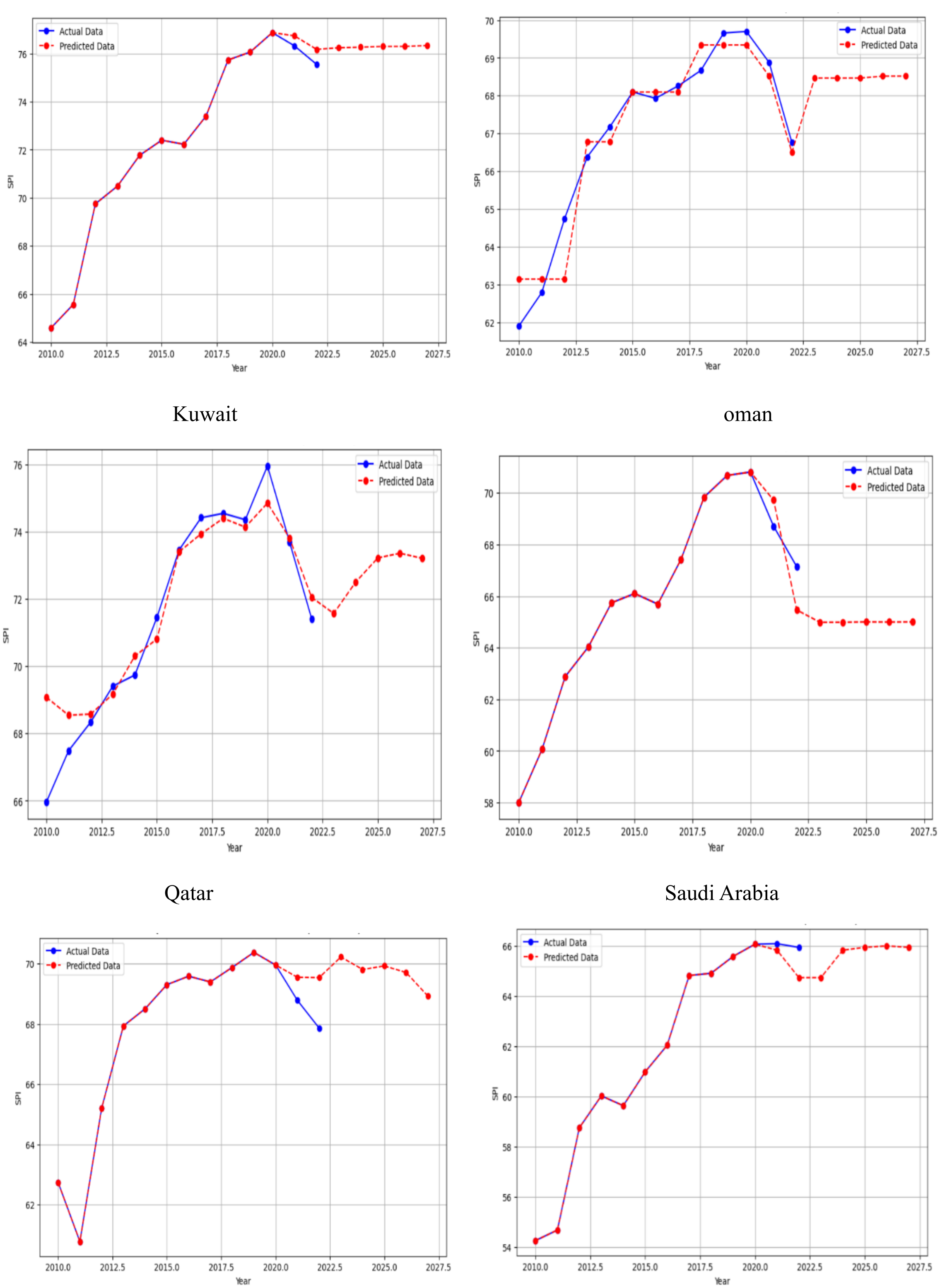

Forecasts show that in the United Arab Emirates, the size of this index will increase for the years 2023 to 2027, but this increase will occur with a very gentle slope. Like the United Arab Emirates, the forecasts show that the country of Bahrain will also undergo significant changes in the area. This is also true for Oman .But according to the forecast made in this research, after the reduction of the social progress index from 2020 to 2022, the changes of this index will increase during the years 2023 to 2027. The forecast results of social progress index in Saudi Arabia also indicate the upward trend of this index.

## Conclusion

The challenge of diversification into knowledge-based economies for the GCC countries region economies, including Oman, Bahrain, Kuwait, the UAE, Qatar, and Saudi Arabia, is a challenge to change. SPI has been one of the major measures adopted in this paper concerning the study of societal well-being and economic development that specify imperatives for ensuring necessary social and economic reforms in the progress of the region. It makes use of data from 2010 to 2023 and machine learning models to indicate that while there has been marked improvement in education, healthcare, and gender equality, personal rights and inclusivity remain highly challenged.

The selected features for six countries are divided into economic, educational, social, environmental, demographic variables subgroups.  These factors determine the social status of GCC countries .Accordingly, the findings of the study indicate that the GCC countries are bound to continue their steady progress beyond 2027, propelled by increased global economic disruption, such as the COVID-19 pandemic. However, full economic diversification requires an ongoing investment in human capital, innovation, and social reforms. It is suggested that policymakers emphasize the need for reduced dependence on hydrocarbons while at the same time encouraging sustainable knowledge-based growth. This is one of the key initiatives that would ensure the social and economic long-term resilience of the region.

Economic diversification must be accompanied by social reforms that ensure no one is left behind. This includes advancing gender equality, ensuring equal access to opportunities regardless of socioeconomic background, and protecting personal rights. Empowering women through education and workforce participation, as well as promoting leadership roles for minorities, will help unlock the full potential of the workforce. Additionally, environmental sustainability must be at the heart of these reforms, focusing on preserving natural resources, promoting green technologies, and reducing pollution. Countries should also work to enhance social safety nets by improving healthcare, housing, and social protection programs to safeguard the well-being of all citizens in times of economic uncertainty or transition.

Thus, on the road to sustainable economic diversification, investment in human capital-both by upgrading structures of education, vocational training, and lifelong learning-should be complemented by fostering innovation through increased R&D and support for entrepreneurship. The logical reduction in dependence on oil should be paralleled by the development of such sectors as renewable energy, technology, and finance. Social reforms have to focus on inclusiveness, gender equality, and personal rights, taken together with environmental sustainability and enhanced social safety nets. It would also ensure the transition toward knowledge-based economies and long-term resilience with increased digital transformation through better infrastructure and cybersecurity, regional cooperation, and international collaboration.